# Electronic instability in pressured black phosphorus under strong magnetic field

Zhong-Yi Wang[a,b], Da-Yong Liu[c], Liang-Jian Zou[a,b,*]

[a] Key Laboratory of Materials Physics, Institute of Solid State Physics, HFIPS, Chinese Academy of Sciences, P. O. Box 1129, Hefei 230031, China

[b] Science Island Branch of Graduate School, University of Science and Technology of China, Hefei 230026, China

[c] Department of Physics, School of Sciences, Nantong University, Nantong 226019, China

**ABSTRACT**

In this paper we have systematically studied the electronic instability of pressured black phosphorous (BP) under strong magnetic field. We first present an effective model Hamiltonian for pressured BP near the *Lifshitz* point. We show that when the magnetic field exceeds a certain critical value, the nodal-line semimetal (NLSM) state of BP with a small band overlap re-enters semiconductive phase by re-opening a small gap. This results in a narrow-band semiconductor with a partially flat valence band edge. We show that above this critical magnetic field, two possible instabilities, i.e., charge density wave (CDW) phase or excitonic insulator (EI) phase, are predicted as the ground state for high and low doping concentrations, respectively. By comparing our results with the experiment, we suggest the field-induced instability observed in recent experiment as EI. Furthermore, we propose that the semimetallic BP under pressure with small band overlaps may provide a good platform to study the magneto-exciton insulators. Our findings bring the first insight into the electronic instability of topological NLSM in the quantum limit.

* Correspondence author, zou@theory.issp.ac.cn

## I. INTRODUCTION

A strong magnetic field can confine the three-dimensional electron gas to its zeroth Landau level, entering the so-called quantum limit (QL) [1]. When condensed matter reaches the QL, it can bring about surprisingly new quantum phenomena, such as fractional quantum Hall effect [2,3] and anomalous negative magnetoresistance [4], and others. Naturally, great attention has been attracted to three-dimensional (3D) novel electronic states in the QL. Within the QL, Landau quantization confines electrons into cyclotron motions with discrete levels in the plane perpendicular to the magnetic field, and the 3D energy spectrum becomes quasi-one-dimensional Landau bands (LBs) dispersing along the field [5]. This leaves the LBs with a very high Landau degeneracy, resulting in an extremely unstable system that tends to form a variety of correlated electron states [6,7]. Exotic phases such as charge-density-wave (CDW) [5,8-11], spin-density-wave (SDW) [11-14], and valley-density-wave [6,15] states, may become as the ground state due to the almost perfect nesting of Fermi surfaces in quasi-one-dimensional spectrum. Furthermore, due to the regulating effect of a magnetic field on the band gap and its enhancement of exciton binding energy [16], a high magnetic field environment is more conducive to the realization of an excitonic insulator (EI) [17-20] in narrow-gap semiconductors and semimetals, as well as Wigner lattice [21-24] and 3D quantum Hall effect (QHE) [6, 25-27].

Since it is very difficult to reach the QL condition of normal metals with high carrier concentration, the experimental observations of the electronic instability have mostly been made in a few low-carrier materials, including celebrated graphite [17,28,29] and bismuth [30-32]. Recently discovered topological materials have aroused great interest in probing instability under high magnetic field due to their small Fermi pocket sizes and band topological nontriviality. From these materials, more novel phenomena have been observed, such as annihilation of Weyl nodes above the QL field [33], helical SDW caused by chiral fermion instability in the Weyl semimetal TaAs [34], and 3D QHEs induced by CDW and unconventional log-periodic quantum oscillations in topological materials $ZrTe_5$ [26,35]. These topological materials provide an important playground for novel quantum phenomena.

Black phosphorus (BP) [36] is a layered semiconductor material with weak van der Waals bonding in the interplanar *z* direction. Xiang *et al.* [37] showed that moderate hydrostatic pressure effectively suppresses its direct band gap and that a *Lifshitz* transition from a semiconductor to a topological semimetal occurs at approximately 1.2 GPa. At the critical pressure, the conduction and valence bands contact at the Z point of Brillouin zone and an anisotropic Dirac semimetal [38] appears with a linear spectrum in the *kx*-axis, i.e., the direction of armchair, and quadratic in the other direction. Around the critical pressure, the conduction and valence bands reverse and BP can host a 3D topological nodal-line semimetal (NLSM) state [39,40]. In further experiment [41], for topological semimetal state of BP at 1.23 GPa, Sun *et al.* observed that an abrupt increase in the out-of-plane magnetoresistance and a kink in the Hall coefficient caused by applying a magnetic field ($\approx 17T$) parallel to the *z*-axis at low-temperature. The abnormal increase in resistance and the temperature dependence of its onset field in the QL demonstrate an electronic phase transition involving many-body effects [42,43]. This is a low-temperature metal-insulator transition behavior induced by a high magnetic field. Such a behavior has been observed in several semimetal materials

[27-29,61], and their mechanisms are regarded as CDW or EI. Some previous theoretical studies [55-57,60] also proposed that applying a magnetic field can induce CDW instability or two competing phases of CDW and EI in Weyl/Dirac systems. However, the instability of topological NLSM in QL has rarely been explored so far.

In this paper we investigate the electronic instability behavior of the CDW and EI in pressured black phosphorus under high magnetic field around *Lifshitz* transition point. Based on the effective model Hamiltonian, we demonstrate that the CDW phase and EI phase are stabilized in different doping ranges in BP, and present low-temperature phase diagrams of the BP around *Lifshitz* transition point in the QL. In the rest of this paper, we begin with an effective model Hamiltonian used to describe BP around *Lifshitz* transition point in Sec. II, and give its Landau level spectrum and the long-range Coulomb interaction in Sec. III. We derive the self-consistent gap equation of CDW and EI in Sec. IV, and discuss the numerically solutions of the gap equation in Sec. V. Finally, we summarize the main results in Sec. VI.

## II. EFFECTIVE MODEL HAMILTONIAN AROUND LIFSHITZ POINT

Although the instability occurs only slightly above the *Lifshitz* transition pressure point, to make our theory more applicable, our model aims to be able to describe the behavior of the region around the transition point. We use the $k \cdot p$ method to describe the BP's electronic structure under hydrostatic pressure. For the ambient bulk BP, the direct energy gap is at the Z point [38]. BP belong to the point group $D_{2h}$ [44] and its orthorhombic (A17) structure can be preserved up to 4.5 GPa [46] under hydrostatic pressure. Thus, based on the analysis of the symmetry [45], the low-energy effective Hamiltonian near the Z point around 1.2 GPa is [40]

$$H = \begin{pmatrix} E_c + a_c k_x^2 + b_c k_y^2 + c_c k_z^2 & \hbar v_f k_x \\ \hbar v_f k_x & E_v - a_v k_x^2 - b_v k_y^2 - c_v k_z^2 \end{pmatrix}, \quad (1)$$

where $E_c$ ($E_v$) is the conduction (valence) band edge energy of bulk BP, $a_{c,v}$, $b_{c,v}$ and $c_{c,v}$ are the respective band parameters, while $v_f$ is the Fermi velocity of the nearly linear band along the armchair (x) direction. And $E_g = E_c - E_v$ is the bandgap (or band overlap for negative values). We discuss our choice of these band parameters around 1.2 GPa below.

Close to the Z point, the band parameters are related to the effective masses via

$$m_{(c,v)x} = \frac{\hbar^2}{2[(\hbar v_f)^2/(E_c - E_v) + a_{(c,v)}]} \ , \ m_{(c,v)y} = \frac{\hbar^2}{2b_{(c,v)}} \ , \ m_{(c,v)z} = \frac{\hbar^2}{2c_{(c,v)}} \ . \quad (2)$$

The band parameters $(a, b, c)_{(c,v)}$ and $v_f$ are chosen such that they yield the known effective masses in the BP of 1.2 GPa. The first-principles calculations in Ref. [38] only gave the effective masses of BP under hydrostatic pressures from 0 to 1 GPa. As shown in Ref. [38], with increase of pressure, only the effective masses $m_x$ decrease significantly, and $m_{y,z}$ almost do not alter. Since $m_{y,z}$ is essentially insensitive to pressure, based on the Eq. (2), we can assume that the pressure has little effect on the band parameters $(a, b, c)_{(c,v)}$. The main consequence of pressure is to reduce the bandgap $E_g$, which leads to a notable change of $m_x$. Hence, the band parameters $(a, b, c)_{(c,v)}$ near 1.2 GPa can be determined by the known effective masses in 1 GPa, i.e., $m_{(c,v)x} = 0.05m_0$, $m_{cy} = 1.21m_0$, $m_{vy} = 0.78m_0$, $m_{cz} = 0.13m_0$, $m_{vz} = 0.31m_0$. As for $v_f$ near 1.2GPa, based on the pressure dependence of average Fermi velocity of BP calculated in Ref. [38], we estimate it to be $2 \times 10^5 m/s$. And it should be noted that the pressure dependence of the energy gap is linear [47] and here $dE_g/dP$ is about 0.23eV/GPa. So, in this work, we vary only the $E_g$ and keep the other band parameters constant.

TABLE I. Band parameters.

| $i$ | $a_i$ | $b_i$ | $c_i$ | $v_f$ |
|---|---|---|---|---|
| $c$ | 41.3 $eV$ Å$^2$ | 3.16 $eV$ Å$^2$ | 29.4 $eV$ Å$^2$ | $2 \times 10^5 m/s$ |
| $v$ | 41.3 $eV$ Å$^2$ | 4.89 $eV$ Å$^2$ | 12.3 $eV$ Å$^2$ | |

To verify the accuracy of the model, we use Eq. (1) to draw the Fermi surface of the semimetal state ($E_g < 0$) of BP in Fig. 1, which is consistent with the results of first principles electronic structure calculations around Z point [38]. This is a compensating semimetal state: the two yellow pockets are hole type, and the others are electron type. And consider that previous studies have reported that BP semimetal is NLSM [39,40], we rewrite Eq. (1) as

$H = H_0 + H'$,

$H_0 = \frac{1}{2}[(a_c+a_v)k_x^2 + (b_c+b_v)k_y^2 + (c_c + c_v)k_z^2 - |E_g|]\sigma_z + \hbar v_f k_x \sigma_x$,

$H' = \frac{1}{2}[E_c + E_v + (a_c-a_v)k_x^2 + (b_c-b_v)k_y^2 + (c_c - c_v)k_z^2]I,$ (3)

where $E_g < 0$, $\sigma_{z,x}$ are Pauli matrices acting on the two band space, $H_0$ describes a conventional NLSM [48] that we know well, with nodal ring satisfying elliptic equation $(b_c + b_v)k_y^2 + (c_c + c_v)k_z^2 = -E_g$ in the $k_y$-o-$k_z$ plane, and $H'$ has no influence on nodal ring equation but causes a variation in the energy of the nodal line. Therefore, $H$ in Eq. (3) is indeed a Hamiltonian of NLSM.

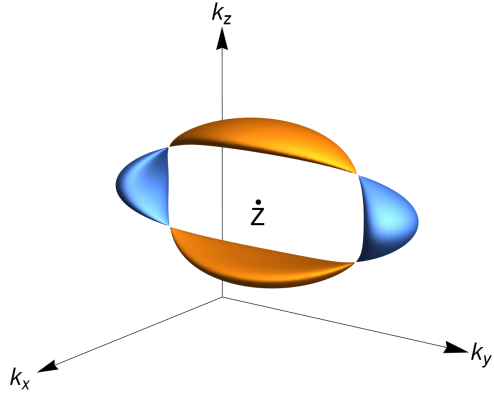

FIG. 1. The Fermi surface of BP semimetallic state under pressure near the Z point drawn based on the $k \cdot p$ model in Eq. (1). The two yellow pockets are hole type, and the others are electron type.

### III. LANDAU LEVEL SPECTRUM AND MANY-BODY INTERACTION

The strong magnetic field is applied along the z direction, i.e., parallel to the plane in which the nodal ring lies. With the Landau gauge, $\vec{A} = -By\vec{x}$, the momentum operator in (1) can be rewritten as $\hbar k_i \to \Pi_i = -i\hbar \partial_i + eA_i$ in terms of the Peierls substitution, where $i = x, y, z$. Obviously we have the commutation relation $[\Pi_x, \Pi_y] = -i\hbar eB$, thus we can define

$$\Pi_x = \frac{\hbar}{\sqrt{2}l_B}(a^+ + a), \quad \Pi_y = \frac{\hbar}{\sqrt{2}l_B i}(a^+ - a), \tag{4}$$

where $l_B = \sqrt{\hbar/eB}$ is the magnetic length. The ladder operator $a$ satisfies $[a, a^+] = 1$. And we get $a^+ a = \frac{1}{2\hbar eB}[\Pi_x^2 + \Pi_y^2] - \frac{1}{2}$ and its eigen-equation $a^+ a |n, k_x> = n|n, k_x>$ with $n=0,1,2\cdots$, where $<y|n, k_x> = \phi_n(y - l_B^2 k_x)$ is a simple harmonic oscillator eigenstate with an oscillation center of $y_0 = l_B^2 k_x$.

Consequently, we arrive at the following Hamiltonian

$$H = \begin{pmatrix} E_c + \frac{a_c(a^+ + a)^2}{2l_B^2} - \frac{b_c(a^+ - a)^2}{2l_B^2} + \frac{c_c \Pi_z^2}{\hbar^2} & \frac{\hbar v_f}{\sqrt{2}l_B}(a^+ + a) \\ \frac{\hbar v_f}{\sqrt{2}l_B}(a^+ + a) & E_v - \frac{a_v(a^+ + a)^2}{2l_B^2} + \frac{b_v(a^+ - a)^2}{2l_B^2} - \frac{c_v \Pi_z^2}{\hbar^2} \end{pmatrix}. \tag{5}$$

Its eigenvalue problem can only be solved numerically. We expand its eigenvector $|\alpha>$ in terms of the harmonic oscillator basis wavefunction $<y|n, k_x>$,

$$\Psi_\alpha(r) = <r|\alpha> = <r|N, k_z, k_x> = \frac{1}{\sqrt{\Omega}} e^{i(k_z z + k_x x)} \varphi_{N,k_z,k_x}(y), \tag{6}$$

where an eigenstate α corresponds to a set of good quantum number $N, k_z, k_x$, $\Omega = L^2$ is the sample size in the $\hat{x}$-o-$\hat{z}$ plane, $\varphi_{N,k_z,k_x}(y) = \begin{pmatrix} \sum_{n=0}^{n_{max}} N_{n,+}(k_z) \phi_n(y - \frac{\hbar}{qB} k_x) \\ \sum_{n=0}^{n_{max}} N_{n,-}(k_z) \phi_n(y - \frac{\hbar}{qB} k_x) \end{pmatrix}$, and $n_{max}$ sets the truncation of the expansion. Note that $N = 0, \pm 1, \ldots$ is Landau energy band index and its corresponding coefficients $N_{n,\pm}(k_z)$ are determined from the energy eigenproblem $H|\alpha> = E_\alpha |\alpha>$. Numerically, we set a large truncation condition $n_{max} = 100$ to ensure convergence of low energy part.

In view of BP upon just entering the semimetal state, $E_g$ is set to a small value -15meV, here. We plot the evolution of the $k \cdot p$ energy spectrum of this semimetallic BP with increasing magnetic field in Fig. 2. As shown in Fig. 2 (a), we can see that the touch points of the conduction and valence bands form a continuous line (nodal line) under magnetic field, which can be explained in terms of a 2D Dirac-cone-pair model [48]. This nodal line can be a flat band when the $H' = 0$ in Eq. (3), as some theoretical studies [48,50-52] have shown. As the applied field increases, from Fig. 2(b) to (d), the closed gap reopens at about 10T. After that, BP becomes a narrow-band semiconductor with partially flat of valence band edge. Such a narrow band gap and a high effective mass at the top of the valence band are probably favorable for the formation of EI phase.

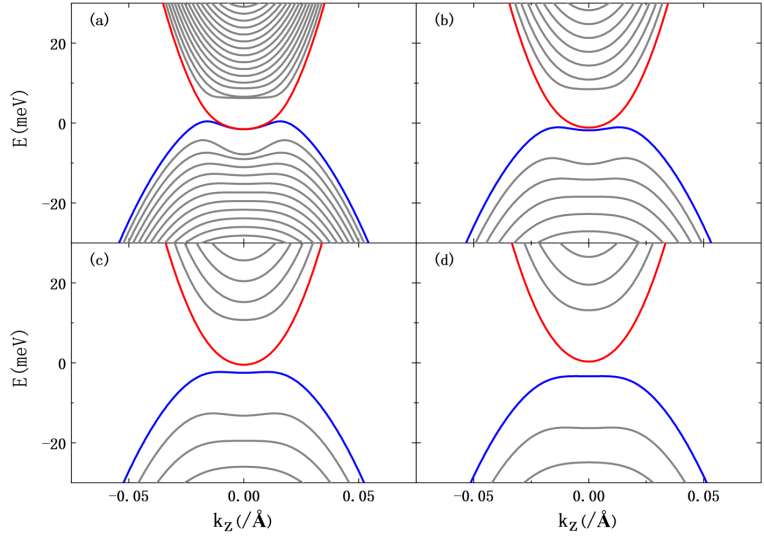

FIG. 2. The Landau bands of semimetal state ($E_g$ =-15meV) of BP at different magnetic field B. (a) B=5T; (b) B=10T; (c) B=15T; (d) B=20T.

Considering the presence of electron-electron interaction and the low carrier concentration of the system, it is appropriate to take the long-range Coulomb repulsive potential into account. It can be written as

$$V = \frac{1}{2}\int dr_1^3 \int dr_2^3 \frac{e^2}{4\pi\varepsilon_0\kappa_0|\vec{r_1}-\vec{r_2}|}\Psi^+(r_1)\Psi^+(r_2)\Psi(r_2)\Psi(r_1), \quad (7)$$

$$\Psi(r) = \sum_\alpha \Psi_\alpha(r) C_\alpha, \quad (8)$$

where $\Psi_\alpha(r)$ is single-particle eigen wavefunctions from (6).

Thus we have

$$V = \frac{1}{2}\frac{e^2}{4\pi\varepsilon_0\kappa_0}\frac{1}{L^3}\sum_{\vec{q}\neq 0}\sum_{k_z,k_x}\sum_{p_z,p_x}\sum_{N,M,N',M'} V(\vec{q}) <N,p_z+q_z,p_x+q_x|e^{iq\cdot r}|M,p_z,p_x><$$

$$N',k_z,k_x|e^{-iq\cdot r}|M',k_z+q_z,k_x+q_x> C^+_{N,p_z+q_z,p_x+q_x} C^+_{N',k_z,k_x} C_{M',k_z+q_z,k_x+q_x} C_{M,p_z,p_x}, \quad (9)$$

where $N,M,N',M'$ are LBs indices, $\vec{q} = (q_x, q_y, q_z)$, $V(\vec{q}) = 4\pi/q^2$ and $\kappa_0 = 10$ is relative dielectric constant [49] of BP. Considering that the system enters the QL, to concentrate on the low-energy physics, we restrict our study to the $N, M, N', M' = c, v$. Here $c$ and $v$ denote the conduction band (the red band in Fig.2) and valence band (the blue band in Fig. 2), respectively. And it should be noted

$$<N',p_z,p_x|e^{-iq\cdot r}|M',k_z,k_x>=$$

$$\delta_{k_z,p_z+q_z}\delta_{k_x,p_x+q_x}\sum_{m,n,\sigma=\pm} N'_{m,\sigma}(p_z)^* M'_{n,\sigma}(k_z) \exp\left(\frac{-il_B^2 q_y(k_x+p_x)}{2}\right) <m,0|D(\frac{l_B}{\sqrt{2}}(-iq_y-$$

$$(p_x-k_x))|n,0>, \quad (10)$$

where $<m,0|D(\xi)|n,0> = (n!/m!)^{1/2}\xi^{m-n}e^{-|\xi|^2/2}L_n^{m-n}(|\xi|^2)$ [53], and $L_n^{m-n}$ is the Laguerre polynomial.

Considering the screening effect between electrons and the enhancement effect of magnetic field on the energy gap between LBs, we take the random-phase approximation (RPA) to derive an effective interaction, $V(\vec{q})$ can be substituted by an effective interaction $V^s(\vec{q}) = V(\vec{q})/[1 +$

$e^2 v(q)\Pi(q,0)/4\pi\varepsilon_0\kappa_0]$, where the bare susceptibility is [54]

$$\Pi(q,\omega) = -\frac{g_s}{2\pi l_B^2}\sum_{N_p,N_p'}\frac{1}{2\pi}\int dk_z \frac{\left(f_{k_z,N_p}-f_{k_z+q_z,N_p'}\right)}{(\hbar\omega+i\eta)+\left(\varepsilon_{k_z,N_p}-\varepsilon_{k_z+q_z,N_p'}\right)}\cdot|<N_p,k_z,k_y|e^{-iq\cdot r}|N_p',k_z+q_z,k_y+q_y>|^2, \quad(11)$$

here $g_s = 2$ is spin degeneracy and $N_p, N_p' = 0, \pm1, ...$ are the LBs indices. One can see that the effective interaction is tunable by magnetic field and temperature.

## IV. ELECTRONIC INSTABILITIES IN STRONG MAGNETIC FIELD

In this section, we use the mean-field (MF) approach to study the electronic instability of pressured BP under strong magnetic field. When a strong field is applied, the spectrum of the 3D electron gas becomes quasi-one-dimensional in the QL regime and the nesting of Fermi surface is naturally satisfied. So the CDW instability is a possible mechanism under a strong magnetic field. In this quasi-one-dimensional case, the CDW wavevector $q_{cdw}$ is $(0,0,2Q)$, and we have order parameter $<C^+_{N,k_z,k_x}C_{M,k_z+q_z,k_x+q_x}>=\eta_{N,M}(k_z)\delta_{q_x,0}\delta_{q_z,2Q}$ in the CDW phase. It is well known that the bare susceptibility peaks at each nesting wavevector and the $q_{cdw}$ is derived from the maximal peaks [55]. As shown in the Landau bands in Fig. 2, it is easy to see that before the gap is opened at B<10 T, there are more than one nesting wavevectors when the doping is low; after the band gap is opened at B>10 T, there is only one nesting wavevector $q_{nest}=(0,0,2k_F)$. For simplicity, we only focus on the state of BP above 10 T. In this case, the $q_{nest}=q_{cdw}$ ($Q=k_F$) and is an intraband nesting vector. Thus we have $<C^+_{N,k_z,k_x}C_{N,k_z+q_z,k_x+q_x}>=\eta_N(k_z)\delta_{q_x,0}\delta_{q_z,2k_F}$ in the CDW phase, where $N$ represents $c$ (or $v$) for electron (or hole) doping.

As mentioned in the Sec. III, with the increase of magnetic field, BP under pressure transits from a semimetal into narrow-band semiconductor with partially flat of valence band edge. Thus, the possibility of an EI transition cannot be ignored. EI is a ground state with the "exciton condensation" and exciton is a spontaneous formation of electron–hole pairs. So we have $<C^+_{N,k_z,k_x}C_{M,k_z+q_z,k_x+q_x}>=\eta(k_z)\delta_{N,\bar{M}}\delta_{q_x,0}\delta_{q_z,0}$ in the EI phase. Here we also only focus on the semiconductor state of BP above 10T because our subsequent results show that the screening effect of the semimetallic BP is strong, the electron-hole interaction becomes weak, hence is unfavorable of the formation of the EI phase.

According to the discussion above, within the framework of the MF theory, we get the following effective interaction terms in the formation of the CDW order and EI order, respectively:

### 1. CDW phase

In the dominant channel of the CDW order, the effective interaction becomes

$$V_{cdw} = -1/2\sum_{k_{z,x}}\sum_{p_{z,x}}V_{p_{z,x}|k_{z,x}}(2k_F)C^+_{N,p_z+k_F,p_x}C^+_{N,-k_F+k_z,k_x}C_{N,k_z+k_F,k_x}C_{N,-k_F+p_z,p_x} + h.c.. \quad(12)$$

We define the CDW order parameter as

$$\Delta^C_{k_{z,x}} = \sum_{p_{z,x}}V_{p_{z,x}|k_{z,x}}(2k_F)<C^+_{N,p_z+k_F,p_x}C_{N,-k_F+p_z,p_x}>. \quad(13)$$

Then we get the MF Hamiltonian

$$H_{MF} = \sum_{k_x,|k_z|<k_F}\begin{pmatrix}C^+_{N,-k_F+k_z,k_x} & C^+_{N,k_F+k_z,k_x}\end{pmatrix}\begin{pmatrix}\varepsilon_N(-k_F+k_z)-\mu_F & -\Delta^C_{k_{z,x}} \\ -\Delta^{C*}_{k_{z,x}} & \varepsilon_N(k_F+k_z)-\mu_F\end{pmatrix}\begin{pmatrix}C_{N,-k_F+k_z,k_x} \\ C_{N,k_F+k_z,k_x}\end{pmatrix}, \quad(14)$$

where $\mu_F$ is Fermi energy and $\varepsilon_N(k_z)$ is energy of conduction band ($N=c$) or valence band ($N=v$).

The self-consistent equations of the CDW order parameter are

$$\Delta^C_{k_{z,x}} = \sum_{p_x,|p_z|<k_F}[f(\omega^C_-(\boldsymbol{p}),T) - f(\omega^C_+(\boldsymbol{p}),T)] V_{p_z,p_x|k_z,k_x}(2k_F) \frac{\Delta^C_{p_{z,x}}}{2E^C_{p_{z,x}}},$$

$$\sum_{p_x,|p_z|<k_F}[f(\omega^C_+(\boldsymbol{p}),T) + f(\omega^C_-(\boldsymbol{p}),T)] = |\Delta n|, \tag{15}$$

where $\Delta n$ is doping concentration, $\omega^C_\pm(\boldsymbol{p}) = -\mu_F + \frac{[\varepsilon_N(-k_F+p_z)+\varepsilon_N(k_F+p_z)]}{2} \pm E^C_{p_{z,x}}$, $E^{C\,2}_{p_{z,x}} = \frac{[\varepsilon_N(-k_F+p_z)-\varepsilon_N(k_F+p_z)]^2}{4} + |\Delta^C_{p_{z,x}}|^2$, $T$ is the temperature and $f(\varepsilon,T)$ is the Fermi-Dirac distribution function. The expression of coupling constants $V_{p_{z,x}|k_{z,x}}(2k_F)$ see in Appendix A.

### 2. EI phase

In the dominant channel of the EI phase, the effective interaction becomes

$$V_{EI} = -\frac{1}{2}\sum_{k_{z,x}}\sum_{p_{z,x}}[V_{p_{z,x}|k_{z,x}}(+)C^+_{c,p_z,p_x}C^+_{v,k_z,k_x}C_{c,k_z,k_x}C_{v,p_z,p_x} +$$

$$V_{p_{z,x}|k_{z,x}}(-)C^+_{c,p_z,p_x}C^+_{c,k_z,k_x}C_{v,k_z,k_x}C_{v,p_z,p_x}] + h.c.. \tag{16}$$

We define the EI order parameter as

$$\Delta^E_{k_{z,x}} = \sum_{p_{z,x}}[V_{p_z,p_x|k_z,k_x}(+) <C^+_{v,p_z,p_x}C_{c,p_z,p_x}> + V_{p_z,p_x|k_z,k_x}(-) <C^+_{c,p_z,p_x}C_{v,p_z,p_x}>]. \tag{17}$$

Then we get the MF Hamiltonian

$$H_{MF} = \sum_{k_{z,x}}\begin{pmatrix}C^+_{c,k_z,k_x} & C^+_{v,k_z,k_x}\end{pmatrix}\begin{pmatrix}\varepsilon_c(k_z)-\mu_F & -\Delta^E_{k_{z,x}} \\ -\Delta^{E\,*}_{k_{z,x}} & \varepsilon_v(k_z)-\mu_F\end{pmatrix}\begin{pmatrix}C_{c,k_z,k_x} \\ C_{v,k_z,k_x}\end{pmatrix}. \tag{18}$$

The self-consistent equations of the EI order parameter are

$$\Delta^E_{k_{z,x}} = \sum_{p_{z,x}}[f(\omega^E_-(\boldsymbol{p}),T) - f(\omega^E_+(\boldsymbol{p}),T)]\left[V_{p_{z,x}|k_{z,x}}(+)\frac{\Delta^E_{p_{z,x}}}{2E^E_{p_{z,x}}} + V_{p_{z,x}|k_{z,x}}(-)\frac{\Delta^{E\,*}_{p_{z,x}}}{2E^E_{p_{z,x}}}\right],$$

$$\sum_{p_{z,x}}\{[f(\omega^E_+(\boldsymbol{p}),T) + f(\omega^E_-(\boldsymbol{p}),T)] - 1\} = \Delta n, \tag{19}$$

where $\omega^E_\pm(\boldsymbol{p}) = -\mu_F + \frac{[\varepsilon_c(p_z)+\varepsilon_v(p_z)]}{2} \pm E^E_{p_{z,x}}$, and $E^{E\,2}_{p_{z,x}} = \frac{[\varepsilon_c(p_z)-\varepsilon_v(p_z)]^2}{4} + |\Delta^E_{p_{z,x}}|^2$. We can see that the value of $\Delta^E_{k_{z,x}}$ is restricted and it is favored for the $\Delta^E_{k_{z,x}}$ to take real values. The expression of coupling constants $V_{p_{z,x}|k_{z,x}}(\pm)$ see in Appendix A.

It should be noted that $V_{p_{z,x}|k_{z,x}}(\pm \text{ or } 2k_F)$ is the function of $(p_x - k_x)$ and $\varepsilon_{c(v)}(k)$ only depends on $k_z$. As a result, we can safely ignore the dependence of order parameter on the x-component momentum $k_x$, i.e., $\Delta^{C(E)}_{k_{z,x}} = \Delta^{C(E)}_{k_z}$. The order parameter $\Delta^{C(E)}_{k_z}$ directly describes the CDW(EI) energy gap.

## V. NUMERICAL RESULTS AND DISCUSSION

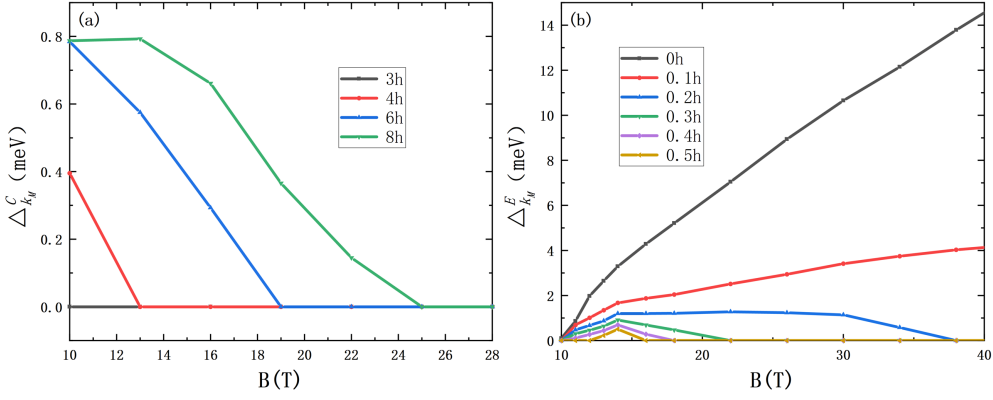

FIG. 4. The maximal value of the CDW (a) and EI (b) orders as functions of magnetic field in T=1K at different concentrations of hole doping. Here the 1h denote $1\times10^{17}cm^{-3}$ hole doping and the band overlap of semimetallic BP is $E_g$=-15meV.

Firstly, we present the magnetic field-dependence of the CDW and EI orders on different doping concentrations in Fig.4. For convenience, $\Delta_{k_M}^{C(E)}$ on the vertical axis represents to the maximal value of the order parameter $\Delta_{k_z}^{C(E)}$ at $k_z = k_M$. For the CDW phase shown in Fig. 4a, we can see that its order fades out as the magnetic field increases. This situation is easy to understand. According to previous research [9,17,57] on CDW under a magnetic field, the strength of CDW order usually has the following relationship: $\Delta \sim k_B T_c \sim E_F \exp(-\frac{1}{N(E_F)u})$, where $E_F$ is the Fermi energy measured from the bottom of the Landau subbands, $N(E_F)$ is the density of states at the $E_F$, and u is related to the electron-electron interaction strength. As the magnetic field increases, the Fermi wave vector [59] $k_F = |\Delta n|\pi^2 l_B^2$ will decrease, and the corresponding $E_F$ will also decrease. The $E_F$ of doping concentration $\Delta n = 6 \times 10^{17} cm^{-3}$ at 20 T is only about 1.5 meV. Such a small $E_F$ is not enough to maintain the existence of the CDW order. So increasing the doping concentration will increase the $E_F$ and then effectively improve the CDW order in the high magnetic field region, which can also be observed in the Fig. 4a. But it should be noted that the doping concentrations cannot be too large, especially in low field region, otherwise the Fermi surface will exceed the QL. Above the QL, due to multiband interactions, the CDW order may become more complicated, which is out of scope of this paper. Within the concentration range that we studied, the typical field-dependent CDW order is no more than 1 meV, so the corresponding maximum transition temperature of the CDW phase is no more than 10 K.

For the EI phase shown in Fig. 4b, in the intrinsic case, the EI order appears after 10 T, i.e., the critical value of band gap reopening, and keeps increasing as the magnetic field increases and easily goes up to about 10 meV at B=30 T. However, for the BP system, there are two points that differ from the original theoretical predictions of a magnetic field-tuned EI phase in graphite [16,29,58]. Firstly, there is no EI order in the region where the energy bands overlap due to strong screening effect between electrons in the BP system. Secondly, the EI order does not reach its maximum value when the band gap just separates. As shown in Eq. (19), when $|\varepsilon_c(p_z) - \varepsilon_v(p_z)|$ reaches its minimum value, it seems to be most conducive to the formation of EI order. But from Eq. (11), one can see that small $|\varepsilon_c(p_z) - \varepsilon_v(p_z)|$ also enhances the screening effect contributed by interband transition, weakens the interaction between electrons, hence be unfavorable to the formation of EI order. Therefore, contrary to the original theoretical prediction, when the band gap further increases, EI order is enhanced instead. At the same time, we can see that doping obviously enhances the screening effect. Even a small amount of doping, for example, $10^{16}$ cm$^{-3}$ orders of magnitude, significantly weakens the EI order and leads to its destruction.

It is worth noting that it is hard to use the relationship $\Delta \sim k_B T_c$ to estimate the transition temperature of EI here. As shown in Fig. 5, we plot the temperature-dependence of the EI orders for intrinsic BP with band gaps 1 meV for B=12 T and 3 meV for B=18 T. It can see that a little change in temperature, for example T=5 K, results in a significantly decrease of several meV in the EI order and the destruction of EI order at B=12 T. This is because the band gap reopened by magnetic field is so narrow that the heat can excite appreciable carriers, effectively enhancing the screening effect.

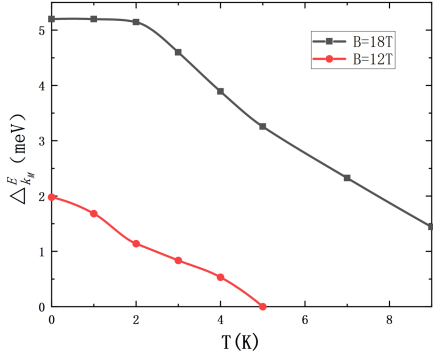

FIG. 5. The maximal value of the EI orders in intrinsic BP under pressure as functions of temperature at B=12 T and 18 T. Here the band overlap of semimetallic BP is $E_g$=-15meV.

We plot the hole-doping density-field phase diagram of topological NLSM BP ($E_g$=-15meV) under pressure in Fig. 6 (b). With no loss of generality, we also consider the cases of $E_g$=-15∓5meV in Fig. 6 (a), (c). We can see that the CDW phase and the EI phase are stable under the QL, and the CDW phase is separated from the EI phase. Compared with the EI phase, the CDW phase could only establish in relatively high-density region. For example, as shown in Fig. 6 (b): within the field range that we studied, the minimum concentration required for the formation of the CDW order is greater than $3.5\times10^{17}$cm$^{-3}$; but the concentration limitation for the formation of EI is $5\times10^{16}$cm$^{-3}$. And from Fig. 6 (a) to (c), a little variation in $E_g$ has no significant effect on CDW phase. But for EI phase, as the degree of band inversion deepens, the onset field of EI phase is moving towards the higher magnetic field and the phase boundary is expanded.

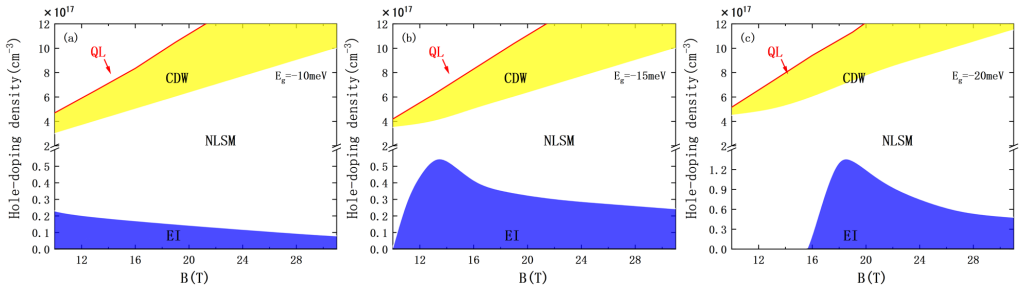

FIG. 6. The phase diagram of pressured BP in hole-doping density-magnetic field parameter space, theoretical parameters are T= 1K, the band overlap of semimetallic BP is (a) $E_g$=-10meV, (b) $E_g$=-15meV, (c) $E_g$=-20meV. Here the red line indicates the critical field of the quantum limit (QL), and the EI, CDW phases are represented by blue, and yellow areas, respectively.

It is interested to compare our present results with recent experimental observation by Sun *et al.* [41]. As stated in Sun *et al.* paper: BP is a hole-doped semiconductor; at 1.23GPa, it enters the compensated semimetallic state; its quantum oscillation frequency of hole-type Fermi surface is 3.6T, and the corresponding cross-sectional area of the Fermi surface is 0.034nm$^{-2}$ [41]. Assuming that the Fermi surface is spherical, we can use the cross-sectional area of the Fermi surface to estimate that the concentration of hole carriers is $n_h=3.4\times10^{16}$cm$^{-3}$, which is less than the order of $10^{17}$. Considering the system in the compensated semimetallic state, the net doping concentration is only going to be less than $n_h$. Therefore, the doping concentration in system is not enough to the formation of the CDW order but suitable to that of the EI order. Moreover, within strong magnetic field range (10~30T) of the experiment [41], our results show that the CDW order only decreases with increasing magnetic field, while the EI order is opposite. As long as the doping concentration is not so high, the EI order will increase with increasing magnetic field, which corresponds to the experimentally observed magnetoresistance anomaly behavior [41]. Thus, we identify the electronic instability observed in the experiment as the EI phase when the pressured BP is compensated or nearly compensated.

Based on above theoretical results, we would like to point out that semimetallic BP around the *Lifshitz* transition point may be a good candidate for the realization of magneto-exciton insulator. The formation of the EI phase requires large exciton binding energies and small band gaps [16,18,19]. Small (or zero) band gaps can be easily achieved in typical topological Dirac/Weyl semimetals, but their linear dispersion still holds even in strong magnetic fields [56,57], which is unfavourable for enhancing exciton binding energy. Since only one direction of the NLSM state in pressured BP is linearly dispersive, i.e., the direction perpendicular to the nodal line plane, when the magnetic field is parallel to the nodal line plane, the linear part of dispersion is destroyed by the Landau quantization while the mass term along the direction of the magnetic field is preserved. This will effectively enhance the exciton binding energy. In addition, as mentioned earlier, a narrow band gap is more conducive to maintain an exciton insulator at finite temperatures than a zero band gap. Thus the BP under pressure with small band overlaps may provide a good platform to study the EI, due to its narrow band gap and relatively large effective mass of electrons and holes under strong field. In fact, due to the strong anisotropy of BP, the effective mass of electrons and holes in the y (zigzag) direction is much larger than that in the z direction. Similarly, when a magnetic field is applied parallel to the nodal plane, it may be easier to obtain EI phase along the y direction than along the z direction. Therefore, further angle-resolved magnetoresistance measurements are strongly encouraged for BP that has just entered the semimetallic state under pressure.

## VI. SUMMARY

In summary, we systematically studied the electronic instability behavior of semimetallic BP under pressure and strong magnetic field. Firstly, based on the effective model Hamiltonian of pressured BP around the *Lifshitz* point, we find that the gap of NLSM state of BP can reopen after the magnetic field is greater than a critical value, and then a narrow-band semiconductor with partially flat of valence band edge appears. With the increase of magnetic field, considering the long-range Coulomb interaction for pressured BP, we have shown the two most possible electronic instabilities, i.e., CDW phase and EI phase, occurring in two different carrier density regimes. We present the low-temperature phase diagrams of the semimetallic BP around *Lifshitz* transition point in the QL and show that CDW phase and EI phase are predicted as the ground state for high and low concentrations of carriers, respectively. By comparing the realistic doping concentration and instability behavior of pressured BP near the *Lifshitz* point, we suggest the field-induced instability observed in recent experiment as EI. Following our theoretical findings, semimetallic BP around the *Lifshitz* transition point may be a good candidate for the realization of magneto-exciton insulator.

ACKNOWLEDGEMENTS

The authors thank the supports from the NSFC of China under Grant nos. 11974354, 11774350 and 11534010, Program of Chinese Academy of Sciences, Science Challenge Project No. TZ2016001. Numerical calculations were partly performed at the Center for Computational Science of CASHIPS, the ScGrid of the Supercomputing Center, the Computer Network Information Center of CAS, and Hefei Advanced Computing Center.

# APPENDIX A: THE EXPRESSION FOR THE COUPLING CONSTANT

The coupling constants $V_{p_{z,x}|k_{z,x}}(2k_F)$ in Eq. (12) are calculated by

$$V_{p_{z,x}|k_{z,x}}(2Q) = \sum_{q_y} \frac{e^2}{4\pi\varepsilon_0\kappa_0} \frac{1}{L^3} [V^s(p_x - k_x, q_y, p_z - k_z) <N, p_z+Q, p_x|e^{iq_y \cdot y}|N, k_z+Q, k_x>$$
$$<N, -Q+k_z, k_x|e^{-iq_y \cdot y}|N, -Q+p_z, p_x>$$
$$-V^s(0, q_y, 2Q)<N, p_z+Q, p_x|e^{iq_y \cdot y}|N, -Q+p_z, p_x><N, -Q+k_z, k_x|e^{-iq_y \cdot y}|M, k_z+Q, k_x>].$$
(A1)

The second term can be ignored in the zero temperature because of the divergence of bare susceptibility.

The coupling constants $V_{p_{z,x}|k_{z,x}}(\pm)$ in Eq. (16) are calculated by

$$V_{p_{z,x}|k_{z,x}}(+) = \sum_{q_y} \frac{e^2}{4\pi\varepsilon_0\kappa_0} \frac{1}{L^3} V^s(p_x - k_x, q_y, p_z - k_z) <c, p_z, p_x|e^{iq_y \cdot y}|c, k_z, k_x><v, k_z, k_x|e^{-iq_y \cdot y}|v, p_z, p_x>,$$

$$V_{p_{z,x}|k_{z,x}}(-) = \sum_{q_y} \frac{e^2}{4\pi\varepsilon_0\kappa_0} \frac{1}{L^3} V^s(p_x - k_x, q_y, p_z - k_z) <c, p_z, p_x|e^{iq_y \cdot y}|v, k_z, k_x><c, k_z, k_x|e^{-iq_y \cdot y}|v, p_z, p_x>.$$
(A2)

$V_{p_{z,x}|k_{z,x}}(+)$ represents that band index of incoming and outgoing momenta is invariant in the scattering process, and $V_{p_{z,x}|k_{z,x}}(-)$ is the opposite. When the effective coupling between the conduction and valence bands is absent, i.e., $v_f = 0$, the second term can be ignored.

# REFERENCES

[1] P. N. Argyres and E. N. Adams, Phys. Rev. 104, 900 (1956).
[2] Perspectives in Quantum Hall Effects: Novel Quantum Liquids in Low-Dimensional Semiconductor Structures, edited by S. D. Sarma and A. Pinczuk (Wiley, New York, 1997).
[3] D. C. Tsui, H. L. Stormer, and A. C. Gossard, Phys. Rev. Lett. 48, 1559 (1982).
[4] B. Wan, F. Schindler, K. Wang, K. Wu, X. G. Wan, T. Neupert and H.-Z. Lu, J. Phys.: Condens. Matter 30, 505501 (2018).
[5] A. H. MacDonald and G. W. Bryant, Phys. Rev. Lett. 58, 515 (1987).
[6] B. I. Halperin, Jpn. J. Appl. Phys. 26, 1913 (1987).
[7] Z. M. Pan and R. Shindou, Phys. Rev. B 100, 165124 (2019).
[8] X.-T. Zhang, and R. Shindou, Phys. Rev. B 95, 205108 (2017).
[9] H. Fukuyama, Solid State Commun. 26, 783 (1978).
[10] D. Yoshioka and H. Fukuyama, J. Phys. Soc. Jpn. 50, 725 (1981).
[11] V. Celli and N. D. Mermin, Phys. Rev. 140, A839 (1965).
[12] K. Takahashi and Y. Takada, Physica B 201, 384 (1994).
[13] Y. Takada and H. Goto, J. Phys.: Condens. Matter 10, 11315 (1998).
[14] H. Yaguchi, J. Phys.: Condens. Matter 21, 344207 (2009).
[15] Z. Tesanovic and B. I. Halperin, Phys. Rev. B 36, 4888 (1987).
[16] E. W. Fenton, Phys. Rev. 170, 816 (1968).
[17] K. Akiba, A. Miyake, H. Yaguchi, A. Matsuo, K. Kindo, and M. Tokunaga, J. Phys. Soc. Jpn. 84, 054709 (2015).
[18] W. Kohn, Phys. Rev. Lett. 19, 439 (1967).
[19] D. Jerome, T. M. Rice and W. Kohn, Phys. Rev. 158, 462 (1967).
[20] Z. M. Pan, X.-T. Zhang and R. Shindou, Phys. Rev. B 98, 205121 (2018).
[21] W. G. Kleppmann and R. J. Elliott, J. Phys. C: Solid State Phys. 8, 2729 (1975).
[22] S.-W. Tsai, D. L. Maslov and L. I. Glazman, Phys. Rev. B 65, 241102(R) (2002).
[23] J. Alicea and L. Balents, Phys. Rev. B 79, 241101(R) (2009).
[24] T. F. Rosenbaum, S. B. Field, D. A. Nelson and P. B. Littlewood, Phys. Rev. Lett. 54, 241 (1985).
[25] M. Koshino and H. Aoki, Phys. Rev. B 67, 195336 (2003).
[26] F. Tang, Y. Ren, P. Wang, R. Zhong, J. Schneeloch, S. A.Yang, K. Yang, P. A. Lee, G. Gu, Z. Qiao, and L. Zhang, Nature (London) 569, 537 (2019).
[27] F. Qin, S. Li, Z. Z. Du, C. M. Wang, W. Q. Zhang, D. P. Yu, H.-Z. Lu and X. C. Xie, Phys. Rev. Lett. 125, 206601 (2020).
[28] B. Fauqué, D. Le Boeuf, B. Vignolle, M. Nardone, C. Proust, and K. Behnia, Phys. Rev. Lett. 110, 266601 (2013).
[29] Z. Zhu, R. D. McDonald, A. Shekhter, B. J. Ramshaw, K. A. Modic, F. F. Balakirev and N. Harrison, Sci. Rep. 7, 1733 (2017).
[30] B. Fauqué, B. Vignolle, C. Proust, J.-P. Issi, K. Behnia, New J. Phys. 11, 113012 (2009).
[31] L. Li, J. G. Checkelsky, Y. S. Hor, C. Uher, A. F. Hebard, R. J. Cava, and N. P. Ong, Science 321, 547 (2008).
[32] Z. Zhu, J. Wang, H. Zuo, B. Fauqué, R. D. McDonald, Y. Fuseya and K. Behnia, Nat. Commun. 8, 15297 (2017).
[33] C.-L. Zhang, S.-Y. Xu, C. M. Wang, Z. Lin, Z. Z. Du, C. Guo, C.-C. Lee, H. Lu, Y. Feng, S.-

M. Huang, G. Chang, C.-H. Hsu, H. Liu, H. Lin, L. Li, C. Zhang, J. Zhang, X.-C. Xie, T. Neupert, M. Z. Hasan, H.-Z. Lu, J. Wang and S. Jia, Nat. Phys. 13, 979 (2017).

[34] C.-L. Zhang, B. Tong, Z. Yuan, Z. Lin, J. Wang, J. Zhang, C.-Y. Xi, Z. Wang, S. Jia, and C. Zhang, Phys. Rev. B 94, 205120 (2016).

[35] H. Wang, H. Liu, Y. Li, Y. Liu, J. Wang, J. Liu, J.-Y. Dai, Y. Wang, L. Li, J. Yan, D. Mandrus, X. C. Xie, and J. Wang, Sci Adv 4, aau5096 (2018).

[36] M. A, Appl Phys A 39, 227 (1986).

[37] Z. J. Xiang, G. J. Ye, C. Shang, B. Lei, N. Z. Wang, K. S. Yang, D. Y. Liu, F. B. Meng, X. G. Luo, L. J. Zou, Z. Sun, Y. Zhang, and X. H. Chen, Phys. Rev. Lett. 115, 186403 (2015).

[38] P.-L. Gong, D.-Y. Liu, K.-S. Yang, Z.-J. Xiang, X.-H. Chen, Z. Zeng, S.-Q. Shen, and L.-J. Zou, Phys. Rev. B 93,195434 (2016).

[39] J. Zhao, R. Yu, H. Weng, and Z. Fang, Phys. Rev. B 94, 195104 (2016).

[40] R. Fei, V. Tran, and L. Yang, Phys. Rev. B 91, 195319 (2015).

[41] Z. Sun, Z. Xiang, Z. Wang, J. Zhang, L. Ma, N. Wang, C. Shang, F. Meng, L. Zou, Y. Zhang, X. H. Chen, Sci. Bull. 63, 1539 (2018).

[42] S. Tanuma, R. Inada, A. Furukawa, O. Takahashi, Y. Iye, and Y. Onuki, in Physics in High Magnetic Fields, edited by S. Chikazumi and N. Miura (Springer-Verlag, Berlin, New York, 1981), p. 316.

[43] Y. Iye, P. M. Tedrow, G. Timp, M. Shayegan, M. S. Dresselhaus, G. Dresselhaus, A. Furukawa, and S. Tanuma, Phys. Rev. B 25, 5478 (1982).

[44] P. Li and I. Appelbaum, Phys. Rev. B 90, 115439 (2014).

[45] G. L. Bir and G. E. Pikus, Symmetry and Strain-Induced Effects in Semiconductors (Wiley, New York, 1974).

[46] M. Okajima, S. Endo, Y. Akahama, and S.-I.Narita, Jpn. J. Appl. Phys. 23, 15 (1984).

[47] Y. Takao and A. Morita, Phys. B+C (Amsterdam) 105, 93(1981).

[48] H. Yao, M. Zhu, L. Jiang and Y. Zheng, J. Phys.: Condens. Matter 30, 285501 (2018).

[49] X. Zhou, W.-K. Lou, D. Zhang, F. Cheng, G. Zhou, and K. Chang, Phys. Rev. B 95, 045408 (2017).

[50] J.-W. Rhim and Y. B. Kim, Phys. Rev. B 92, 045126 (2015).

[51] L. Oroszlány, B. Dóra, J. Cserti, and A. Cortijo, Phys. Rev. B 97, 205107 (2018).

[52] Y. Wang, T. Boemerich, J. Park, H. F. Legg, A. A. Taskin, A. Rosch, Y. Ando, arXiv:2208.10314.

[53] K. E. Cahill and R. J. Glauber, Phys. Rev. 177, 1857 (1969).

[54] J. Hofmann, Phys. Rev. B 100, 245140 (2019).

[55] M. Trescher, E. J. Bergholtz, M. Udagawa, and J. Knolle, Phys. Rev. B 96, 201101 (2017).

[56] Z. Song, Z. Fang, and X. Dai, Phys. Rev. B 96, 235104 (2017).

[57] K.-Y. Yang, Y.-M. Lu, and Y. Ran, Phys. Rev. B 84, 075129 (2011).

[58] A. A. Abrikosov, Zh. Eksp. Teor. Fiz. 65, 1508 (1973).

[59] P.-L. Zhao, H.-Z. Lu, and X. C. Xie, Phys. Rev. Lett. 127, 046602 (2021).

[60] R.-X. Zhang, J. A. Hutasoit, Y. Sun, B. Yan, C. Xu, and C.-X. Liu, Phys. Rev. B 93, 041108 (2016).

[61] R. Yamada, J. Fujioka, M. Kawamura, S. Sakai, M. Hirayama, R. Arita, T. Okawa, D. Hashizume, T. Sato, F. Kagawa, R. Kurihara, M. Tokunaga and Y. Tokura, Npj. Quantum Mater. 7, 13 (2022).